# Improving Retrieval Results with discipline-specific Query Expansion

Thomas Lüke, Philipp Schaer, Philipp Mayr

GESIS – Leibniz Institute for the Social Sciences,
Unter Sachsenhausen 6-8, 50667 Cologne, Germany
{thomas.lueke|philipp.schaer|philipp.mayr}@gesis.org

**Abstract.** Choosing the right terms to describe an information need is becoming more difficult as the amount of available information increases. Search-Term-Recommendation (STR) systems can help to overcome these problems. This paper evaluates the benefits that may be gained from the use of STRs in Query Expansion (QE). We create 17 STRs, 16 based on specific disciplines and one giving general recommendations, and compare the retrieval performance of these STRs. The main findings are: (1) QE with specific STRs leads to significantly better results than QE with a general STR, (2) QE with specific STRs selected by a heuristic mechanism of topic classification leads to better results than the general STR, however (3) selecting the best matching specific STR in an automatic way is a major challenge of this process.

## 1  Introduction

Users of scientific digital libraries have to deal with a constantly growing amount of accessible information. The challenge of expressing one's information need through the right terms has been described as "vocabulary problem" by Furnas et al. [1]. Specialized knowledge organization systems (KOS) like thesauri or classifications have been created to support the users in the search process and to provide a consistent way of expressing the information need. As a way to provide easy access to these tools methods like Entry Vocabulary Modules [6] or so-called Search-Term-Recommenders (STR) have been introduced. Such recommendation services are also in use in commercial end-user systems like Amazon or eBay.

A common approach in IR systems to improve retrieval results is the use of term based query expansion (QE). If semantically close concepts are added to an initial query term it often shows query results improvements. Typically language and terminology becomes more specific if a terminology in a subject discipline e.g. the social sciences is differentiated into sub disciplines. This has also been proven for suggestions given by STRs [7], however, effects on the application of QE have not been evaluated. Using a heuristic mechanism, our approach utilizes the phenomenon of language specialization to recommend the most specific concepts from a controlled vocabulary through discipline-specific STRs. We conduct an empirical analysis in the domain of social sciences to evaluate retrieval performance in a standard IR evaluation environment using QE.

The paper is structured as follows: Section 2 gives a brief overview of previous findings in the area of search term recommendation, query expansion and IR evaluation. Section 3 describes the methods used in this paper to evaluate our setup. In Section 4 we present and discuss our evaluation. Section 5 summarizes the findings of this paper and presents ideas for future work.

## 2   Related Work

Hargittai [2] has shown that users need supporting mechanisms while expressing their information need through search queries. Such support may be provided by query recommendation mechanisms, which try to enrich the existing query with additional terms. This leads to better retrieval results or provides the searcher with a new viewpoint on the search as shown by Mutschke et al. [5].

Automatic Query Expansion mechanisms have been divided into two classes by Xu and Croft [8]: Expansion recommendations based on a global analysis of the entire document collection and recommendations based on the local subset of documents that were retrieved by using the unexpanded query (called pseudo-relevance feedback [4]). In their research the local approach outperformed the global one clearly. However, when the amount of non-relevant documents in the results to the original query increases, a so-called query drift may occur. Documents not relevant to the users' information need lead to mostly irrelevant expansion terms. If the query gets expanded with such terms, it drifts away from the original meaning and results in even less relevant documents. Mitra et. al [4] have proposed techniques to overcome query drift.

Additionally Petras [7] found that discipline-specific search term recommenders which are trained on sub disciplinary document corpora deliver more specific search term suggestions than general recommenders which are trained on an entire database.

## 3   Discipline-specific Term Recommendation

In our approach we apply Petras' [7] idea of more specific STRs onto an an automatic query expansion setup. According to the findings in [4, 8] we expect retrieval results to improve. With real-world applications in mind and the intention to reduce the chance of query drift through non-relevant data sets we demand STRs to be created a-priori rather than on-demand as with pseudo-relevance feedback. To create document sets belonging to specific disciplines we use a hierarchically structured classification system. It is called "classification of social sciences" and part of the SOLIS[1] data set. For example any class starting with 1 is connected to the entire field of *social scienc-*

---

[1] http://www.gesis.org/en/services/research/solis-social-science-literature-information-system/

Table 1: Overview of the created discipline-specific (DS) STRs. Class describes the specific sub-discipline of the social sciences the STR was based on. #Docs and #CT shows the number of documents or controlled vocabulary terms in that given collection.

| STR | Class | #Docs | #CT | STR | Class | #Docs | #CT |
|---|---|---|---|---|---|---|---|
| DS-1 | Basic Research | 26817 | 5642 | DS-9 | Economics | 45217 | 6213 |
| DS-2 | Sociology | 76342 | 7184 | DS-10 | Social Policy | 26289 | 5586 |
| DS-3 | Demography | 26298 | 5322 | DS-11 | Employment Research | 61742 | 5610 |
| DS-4 | Ethnology | 5409 | 3787 | DS-12 | Women's Studies | 18116 | 5301 |
| DS-5 | Political Science | 95536 | 6995 | DS-13 | Interdisciplinary Fields | 38985 | 6454 |
| DS-6 | Education | 18820 | 5199 | DS-14 | Humanities | 53863 | 6703 |
| DS-7 | Psychology | 24785 | 5725 | DS-15 | Legal Science | 16330 | 5549 |
| DS-8 | Communications | 37285 | 5893 | DS-16 | Natural Science | 6083 | 4015 |
| Global | Social sciences | 383000 | 7781 | | | | |

*es*, any class starting with 102 is connected to the sub-discipline of *sociology* and class 10209 is assigned to documents from the special field of *family sociology*. This structure allowed us to create STRs at the top level that cover every area of the classification system thereby allowing us to choose matching discipline-specific STRs for queries from various disciplines.

To test the effects of recommendation terms from different disciplines we use a standard IR evaluation environment. A set of pre-defined topics, each consisting of a title and a set of documents relevant to that query, is processed on the SOLR[2] search platform which uses *tf-idf* to rank results. The data sets that represent the basis for the different STRs are all created from the SOLIS dataset, a collection of more than 400,000 documents from various disciplines of the social sciences.

Based on the assumption of a specialized vocabulary in different scientific disciplines, 16 custom data sets (see DS-1 to DS-16 in Table 1) for different sub-disciplines of SOLIS are created. These 16 sets as well as the entire SOLIS data set (see Global in Table 1) are the basis for 17 STRs.

In addition SOLIS is indexed with the thesaurus TheSoz[3] which consists of almost 7800 descriptor terms. Our discipline-specific STRs reduce the vocabulary of TheSoz to about 5700 terms per data set (on average) with a trend of even smaller (and presumably more specific) vocabularies. Exact numbers can be found in Table 1.

Each STR is created to match arbitrary input terms to terms of the controlled vocabulary TheSoz. All documents of a data set are processed using a co-occurrence analysis of input terms (found in title and abstract of each document) and the subject-specific descriptor terms assigned to a document. In order to rank the suggested terms from the controlled vocabulary the logarithmically modified Jaccard similarity measure is used.

---

[2] http://lucene.apache.org/solr/
[3] http://thedatahub.org/dataset/gesis-thesoz

## 4   Evaluation

This section contains a description of our evaluation. In the first paragraph the setup is described. The second paragraph presents the main results and effects of discipline-specific expansion on retrieval performance measured through average precision values. In addition we give a single example of a query, going from the broader level to a detailed in-depth inspection of discipline-specific QE.

### 4.1   Evaluation Setup

In order to test the effects of discipline-specific STRs and a general STR within a query expansion scenario we choose the GIRT4 corpus [3], which is a subset of SOLIS. It is used in evaluation campaigns like CLEF or TREC. We use 100 of the CLEF topics ranging from years 2003 to 2006 (topic numbers were 76 to 175) and classify them through a heuristic approach based on the classification system mentioned above: All relevant documents for a given topic are put into groups based on their classification ID. The classification ID of the group that holds the most documents is assumed to be the topic's classification. Queries from these topics are created by removing stop words from the title of each topic. To test the performance of QE we expand the query with three different STRs:

- the general STR, based on the entire SOLIS data set (our baseline)
- the STR of the topic's class
- the STR that performs best for a given topic (out of the 16 discipline-specific STRs)

Every QE is performed automatically with the top 4 recommendations of a STR. We report mean average precision (MAP), rPrecision as well as p@5, p@10, p@20 and p@30. As additional comparison we include the results of a standard installation of SOLR without QE. Every query, whether expanded or not, is processed by this platform. We use Student's t-test to verify significance of the improvements.

### 4.2   Results

Table 2 shows the results of the QE with different STRs. The first observation which can be made is that the discipline-specific STRs always perform better than a general STR (and thereby also improve retrieval performance compared to an unexpanded query). The last line of Table 2 shows the maximum performance possible through the use of our 16 discipline-specific STRs. It is significantly better than a general STR in every case. However, the improvements for those STRs that are chosen based on the classification of topics did not always reach significance. Only the precision within the first 5 and 10 top ranked documents is significantly higher. Still a more precise classification of the original query is necessary to gain maximum benefit from our discipline-specific STRs. In addition to measuring the impact on average precision we further analyze the use of discipline-specific STRs by examining an individual query. According to Petras [6] the results of a QE could significantly improve if the "right" terms are added to the query. We will see how different recommendations influence the results. Topic no. 131 has the title (and thus the query) "bilingual education". Table 3 shows the top 4 recommendations of each STR for this topic. While "multilingualism" is always a recommendation and "child" and "speech" appear twice, the rest of the terms appear only in one recommender. The general recommender proposes the most common terms while the two discipline-specific STRs propose more specific terms. In Table 4 we can see the effects of these different recommendations on retrieval precision. The unexpanded query performs satisfying but leaves room for improvement as it presents only 2 relevant documents within the top 5 and 3 within the top 10 documents (see p@5 and p@10). Expanding the query with terms from the general STR improves these results to 3 and 6 relevant documents in the top 5 or top 10 respectively. Using terms from the pre-defined, topic-fitting, discipline-specific STR 4 out of 5 documents within the top 5 are relevant. Finally, the best performing discipline-specific STR manages to expand the query in a way that all top 10 documents are relevant and even within the top 20 documents only 3 are not relevant.

Table 2 : Evaluation results averaged over 100 topics for three different types of QE. Recommendations are based on a general STR (gSTR) which served as a baseline, a discipline-specific STR fitting the class of the topic (tSTR) and the discipline-specific STR performing best on each topic (bSTR). Confidence levels of significance are: * α = .05, ** α = .01.

| Exp. Type | MAP | rPrecison | p@5 | p@10 | p@20 | p@30 |
|---|---|---|---|---|---|---|
| gSTR (Base) | 0.155 | 0.221 | 0.548 | 0.509 | 0.449 | 0.420 |
| tSTR | 0.159 | 0.224 | 0.578* | 0.542** | 0.460 | 0.424 |
| bSTR | 0.179** | 0.233** | 0.658** | 0.601** | 0.512** | 0.463** |

Table 3: Top 4 recommendations for the input terms "bilingual education" from three STRs

| Recommendation | General | Topic-fitting | Best-performing |
|---|---|---|---|
| 1 | Multilingualism | Child | Multilingualism |
| 2 | Child | School | Speech |
| 3 | Speech | Multilingualism | Ethnic Group |
| 4 | Intercultural Education | Germany | Minority |

Table 4: Evaluation results for topic 131 with three different types of QE Recommendations. Bold font indicates improvement.

| Exp. Type | AP | rPrecison | p@5 | p@10 | p@20 | p@30 |
|---|---|---|---|---|---|---|
| Solr | 0.039 | 0.127 | 0.4 | 0.3 | 0.2 | 0.133 |
| gSTR | **0.072** | **0.144** | **0.6** | **0.6** | **0.4** | **0.333** |
| tSTR | **0.076** | **0.161** | **0.8** | 0.6 | **0.45** | 0.333 |
| bSTR | **0.147** | 0.161 | **1** | **1** | **0.85** | **0.567** |

## 5      Conclusion and Future Work

Our research shows that the use of discipline-specific Search-Term-Recommenders can improve the retrieval performance significantly if used as basis for an automated query expansion. However, it also becomes clear that choosing the best STR in an automated setting of query expansion is far from trivial. By doing an in-depth analysis of a single query we additionally demonstrate how discipline-specific term recommendations can improve the quality of search results for a user. This leads us to the conclusion that discipline-specific STRs can be a valuable addition to expert search platforms where users might not know how to optimally express their search.

In conclusion, STRs that are meant to assist users should be discipline-specific in order to recommend more specific terms. Still, it has to be determined how specific (or small) a data set may be while still producing reasonable results. To improve quality of QE it is essential to have a good algorithm for determining the specific discipline of the query. Besides having more specific recommendation another aspect of further research could be the use of additional metadata fields as it is common for users to enrich their search by explicitly specifying authors or other metadata fields (for further research on recommendations based on different types of metadata see the work by Schaer et al. in these proceedings). A STR providing recommendations of this kind could add additional benefits to a user's search, especially if it recommends e.g. the main authors of a specific discipline.

**Acknowledgements.** This work was partly funded by DFG, grant no. SU 647/5-2.